\documentclass[
  ,draft            
  ,numberedheadings 
  ]
  {aipproc}

\layoutstyle{6x9}

\begin{document}

\title{Low energy nuclear scattering and sub-threshold spectra
from a multi-channel algebraic scattering theory.}



\author{K. Amos}{
  address={School of Physics, University of Melbourne, Victoria 3010,
Australia}
}

\author{P. Fraser}{
  address={School of Physics, University of Melbourne, Victoria 3010,
Australia}
}

\author{S. Karataglidis}{
  address={School of Physics, University of Melbourne, Victoria 3010,
Australia}
}

\author{D. van der Knijff}{
  address={Advanced Research Computing, 
University of Melbourne, Victoria 3010, Australia}
}

\author{J. P. Svenne}{
  address={Department of Physics and Astronomy,
University of Manitoba,
and Winnipeg Institute for Theoretical Physics,
Winnipeg, Manitoba, Canada R3T 2N2}
}

\author{L. Canton}{
  address={Istituto Nazionale di Fisica Nucleare, sezione di Padova, \\
e Dipartimento di Fisica dell'Universit$\grave {\rm a}$
di Padova, via Marzolo 8, Padova I-35131,
Italia}
}

\author{G. Pisent}{
  address={Istituto Nazionale di Fisica Nucleare, sezione di Padova, \\
e Dipartimento di Fisica dell'Universit$\grave {\rm a}$
di Padova, via Marzolo 8, Padova I-35131,
Italia}
}

\begin{abstract}
 A multi-channel algebraic scattering theory, to find solutions of
 coupled-channel scattering problems with interactions determined by
 collective models, has been structured to ensure that the Pauli
 principle is not violated. Positive (scattering) and negative
 (sub-threshold) solutions can be found to predict both the compound
 nucleus sub-threshold spectrum and all resonances due to coupled
 channel effects that occur on a smooth energy varying background.
\end{abstract}

\maketitle


\section{Introduction} 
     Low energy cross sections from the collision of nucleons with light mass 
nuclei show sharp as well as broad resonances upon a smooth, energy dependent 
background. Those resonances may correlate to states in the discrete spectrum 
of the target.    To interpret such scattering data requires use of a complex 
coupled-channel reaction theory. We have developed such a theory~\cite{Am03}; 
one that has very important improvements over those used heretofore.     This
theory, a multi-channel algebraic scattering (MCAS) theory, finds solution of 
the 
coupled Lippmann-Schwinger equations for the scattering of quantal systems in
momentum space, and is a quite general one. To date, however, we have limited 
study to nucleon scattering from targets of zero ground state spin.

The prime information sought are   scattering ($S$) matrices which are easily
extracted from the $T$-matrices generated by MCAS.      The approach involves 
using matrix algebra on matrices built using Sturmian-state expansions of the 
relevant nucleon-nucleus potential matrix.   With this method, all resonances 
in any energy range, can be identified       and their centroids, widths, and 
spin-parities determined. Similarly the energies and spin-parities of   bound 
states of the compound system sub-threshold can be determined.

               It has long been known that collective model prescriptions  of 
nucleon-nucleus scattering  violate the Pauli principle.   It seems that this   
has been the case in all such  calculations made prior to those reported   in 
Ref.~\cite{Am03}.              However, the MCAS procedure enables use of  an 
orthogonalizing pseudo-potential (OPP) approximation   by   which such  Pauli 
principle  violation can be alleviated.    Doing so is crucial to finding the 
parameter values specifying the interaction that  simultaneously   gives  the
sub-threshold  compound nucleus spectrum and the low energy scattering  cross 
sections.

The MCAS approach that is based upon using Sturmian functions as a basis set, 
is outlined next along with the process by  which           resonances can be 
identified and located. Results of calculations made using a collective model 
prescription for the interaction potential matrix are then discussed. In that 
collective model, the interaction field is allowed  to  be deformed      from 
sphericity.  That deformation has been taken to second order in   the case of 
neutron scattering from $^{12}$C allowing coupling to the     ground $0^+_1$, 
$2^+_1$ (4.4389 MeV), and $0^+_2$ (7.64 MeV) states. Results have been  found 
for nucleon energies to 6 MeV. Some new results for the ${}^{6}$He+$p$ system 
are reported as well.

   The theory also allows formation of the optical potential by appropriately
subsuming  the  coupled  channels  equations  into  effective elastic channel 
scattering equations.     The optical potentials thus constructed, as well as
allowing for the Pauli principle, are very non-local and energy dependent. We 
present typical results for the $n+{}^{12}$C case.

\section{The MCAS theory (in brief)}
\label{multiT}

    For a system of $\Gamma$ channels for each allowed scattering spin-parity
$J^\pi$ let the index $c\ (=1,\Gamma)$ denote the set of quantum numbers that 
identify each channel uniquely.    Let $c = 1$ designate the elastic channel.
The integral equation approach in momentum space   for   potential   matrices 
$V_{cc'}^{J^\pi}(p,q)$, requires solution of coupled  Lippmann-Schwinger (LS)
equations giving a multichannel $T$-matrix of the form
\begin{eqnarray}
T_{cc'}(p,q;E) &=& V_{cc'}(p,q)\, 
+\, \frac{2\mu}{\hbar^2} 
\left[ \sum_{c'' = 1}^{\rm open} \int_0^\infty V_{cc''}(p,x) 
\frac{x^2}{k^2_{c''} - x^2 + i\epsilon} T_{c''c'}(x,q;E)
\ dx \right.
\nonumber\\
&&\hspace*{2.8cm}
\left.- \sum_{c'' = 1}^{\rm closed} \int_0^\infty 
V_{cc''}(p,x) \frac{x^2}{h^2_{c''} + x^2} 
T_{c''c'}(x,q;E) \ dx \right],\;
\label{multiTeq}
\end{eqnarray}
where the conserved values of $J^\pi$ is to be understood.       The open and
closed channels contributions have channel wave numbers 
$k_c = \sqrt{\frac{2\mu}{\hbar^2}(E - \epsilon_c)}$ and 
$h_c = \sqrt{\frac{2\mu}{\hbar^2}(\epsilon_c - E)}$,
for $E > \epsilon_c$ and $E < \epsilon_c$ respectively.       $\epsilon_c$ is
the threshold energy of channel $c$ and  $\mu$ is the reduced mass. Solutions
of Eq.~(\ref{multiTeq}) are sought using (finite rank) separable   expansions
of the potential matrix elements, 
$V_{cc'}(p,q) \sim  \sum^N_{n = 1} \chi_{cn}(p)\ \eta^{-1}_n\ \chi_{c'n}(q)$.

     In the MCAS method, the Sturmians that lead to specification of the form 
factors $\chi_{cn}(p)$, are solutions of Schr\"odinger equations with     the
chosen input matrix of potentials.   In coordinate space, if those potentials
are defined from a collective model prescription of the nucleon-target system 
with local forms $V_{cc'}(r)$,  the Pauli principle can be satisfied by using
an OPP  method in the determination of the Sturmians~\cite{Am03,Ca05}.

The link between the multichannel $T$- and $S$-matrices is
\begin{equation}
S_{cc'} = \delta_{cc'} - i^{l_{c'} - l_c +1} \pi \frac{2\mu}{\hbar^2} 
\sum_{n,n' = 1}^N \sqrt{k_c} \chi_{cn}(k_c) \left([\mbox{\boldmath $\eta$} 
- {\bf G}_0]^{-1} \right)_{nn'} \ \chi_{c'n'}(k_{c'})\sqrt{k_{c'}}\ ,
\label{multiS}
\end{equation}
where now $c,c'$ refer to open channels only.         In this representation, 
\mbox{\boldmath $\eta$} has matrix elements      $\left( \eta \right)_{nn'} = 
\eta_n\ \delta_{nn'}$      while      those     of   \textbf{${\bf G}_0$} are 
$\left( {G}_0 \right)_{nn'} = (G)$,
\begin{equation}
(G) = \frac{2\mu}{\hbar^2}\left[ \sum_{c = 1}^{\rm open} 
\int_0^\infty \chi_{cn}(x) \frac{x^2}{k_c^2 - x^2 + i\epsilon} 
\chi_{cn'}(x)\ dx - \sum_{c = 1}^{\rm closed} \int_0^\infty 
\chi_{cn}(x) \frac{x^2}{h_c^2 + x^2} \chi_{cn'}(x)\ dx \right],
\label{xiGels}
\end{equation}
Bound states of the compound system are defined by the zeros of the    matrix
determinant when the energy is $E < 0$. They link to zeros of 
$\{ \left| \mbox{\boldmath $\eta$}-{\bf G}_0\right| \}$
when all channels in Eq.~(\ref{xiGels}) are closed.        For details of the
Sturmian expansions and their use, see Ref.~\cite{Am03}.

An essential feature of the MCAS approach, of importance in studies when very
narrow    resonances    are    to    be    observed, is the resonance finding 
scheme~\cite{Am03}.           Essentially that requires recasting the elastic 
scattering $S$-matrix (for each $J^\pi$) as
\begin{equation}
S_{11} = 1 - i \pi \frac{2\mu}{\hbar^2} \sum_{nn'=1}^M k \ \chi_{1n}(k) 
\frac{1}{\sqrt{\eta_n}} \left[\left({\bf 1} - 
\mbox{\boldmath $\eta$}^{-\frac{1}{2}}
{\bf G}_0\mbox{\boldmath $\eta$}^{-\frac{1}{2}}    
\right)^{-1}\right]_{nn'} \frac{1}{\sqrt{\eta_{n'}}} \chi_{1n'}(k)\, .
\end{equation}
Here, the elements of the diagonal (complex) matrix 
\mbox{\boldmath $\eta$}$^{-\frac{1}{2}}$ are
$\frac{1}{\sqrt{\eta_n}}\delta_{nn'}$.    Then the  complex-symmetric matrix, 
\mbox{\boldmath $\eta$}$^{-\frac{1}{2}}{\bf G}_0$ 
\mbox{\boldmath $\eta$}$^{-\frac{1}{2}}$   can    be diagonalized to find the 
evolution of its complex eigenvalues, $\zeta_r$,      with respect to energy. 
Resonant behavior occurs when one of the complex $\zeta_r$ eigenvalues passes 
close to the point (1,0) in the Gauss plane. The complex part of that   limit
eigenvalue relates to the width of the resonance.         The elastic channel
$S$-matrix has a  pole  structure  at  the  corresponding energy where one of 
these eigenvalues approach unity.

\section{Results using MCAS for $n+A$ systems}

    Results for the scattering of neutrons to $\sim$6 MeV from ${}^{12}$C are
displayed in Fig.~1. Therein the elastic scattering cross section is compared
with data with respect to the ground state of ${}^{12}$C as zero energy.   On
the same scale, in the right hand parts of 
\begin{minipage}[t]{7.0cm}
\vspace*{2mm}
\scalebox{0.3}{\includegraphics*{AmosK-fig1.eps}}
\vspace*{2mm}
\end{minipage} \hfill
\begin{minipage}[t]{7.0cm}
the figure,     the experimental and theoretical sub-threshold  and resonance 
states in ${}^{13}$C are compared.                  We have noted in a recent 
publication~\cite{Ca05} how crucial it is to account  for the Pauli principle 
via the OPP scheme to achieve these results.
\vspace*{5mm}

{\footnotesize {\bf FIGURE 1.} Spectra of ${}^{12,13}$C   and   the   elastic 
cross section (barn) for the $n+{}^{12}$C system.  The labels with the states 
designate values of $2J$ and parity.}
\vspace*{2mm}
\end{minipage} \hfill

In Fig.~2., the elastic cross section for neutron scattering  is shown in the
top panel 
\begin{minipage}[t]{7.0cm}
\vspace*{2mm}
\scalebox{0.38}{\includegraphics*{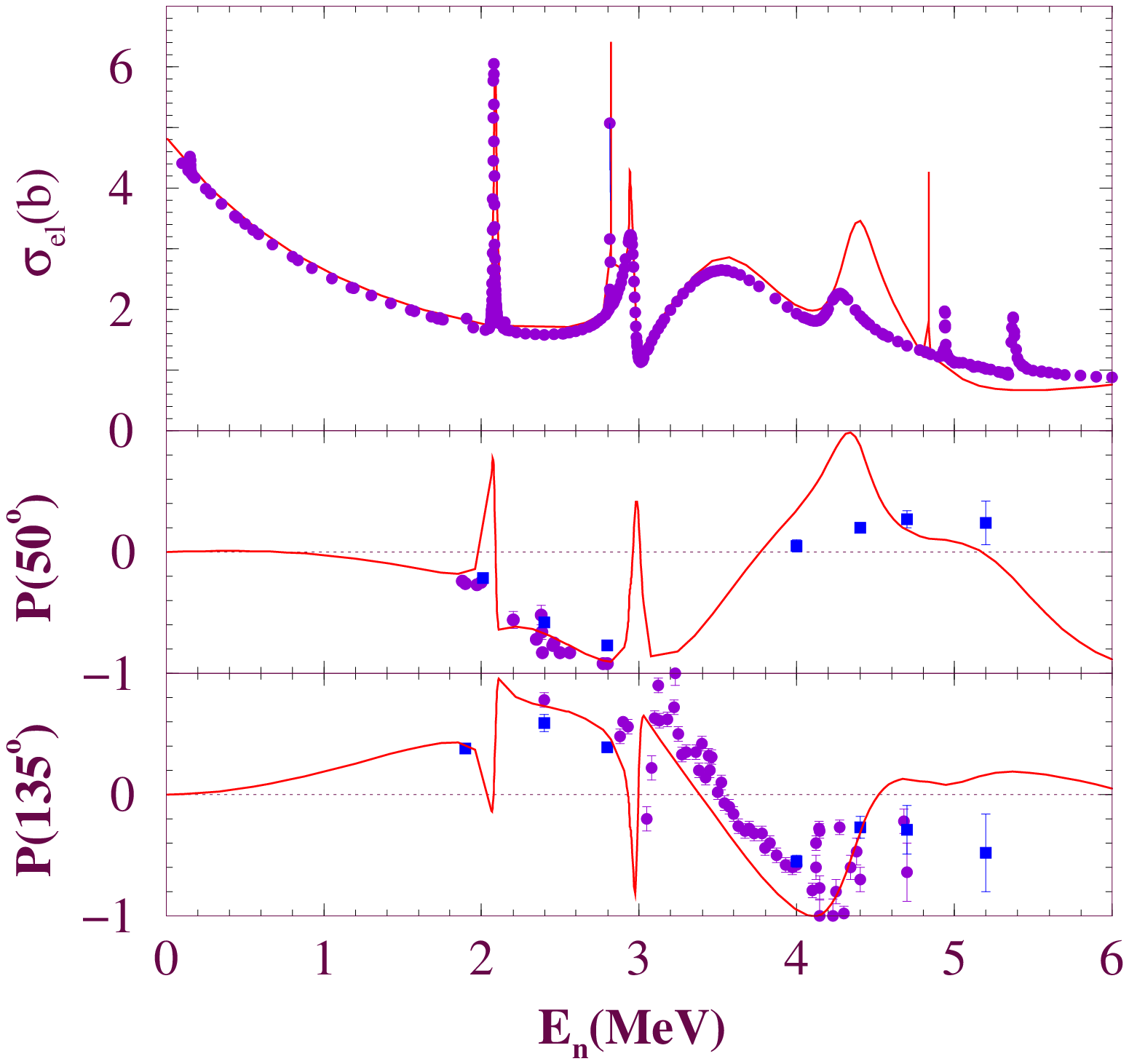}}
\vspace*{2mm}
\end{minipage} \hfill
\begin{minipage}[t]{7.0cm}
while polarizations at two scattering angles are shown underneath.        The 
resonance structures most evident in the cross section are well reproduced by
MCAS and those details are confirmed in the polarizations.                The 
spin-parities of the resonances also coincide with those of known states   in
the spectrum of ${}^{13}$N.  Notably the

\vspace*{5mm}
{\footnotesize {\bf FIGURE 2.}
The elastic cross section for n-${}^{12}$C scattering (top) and polarizations
at 50$^\circ$ (middle) and at 135$^\circ$ (bottom) compared to MCAS results.}
\vspace*{2mm}
\end{minipage}\hfill
prominent $\frac{5}{2}^+$ resonance at 2 MeV,     and the two $\frac{3}{2}^+$ 
resonances spanning 2.8 - 4.0 MeV, are found with very good widths,    narrow
and broad respectively.
\begin{minipage}[t]{7.0cm}
\vspace*{2mm}
\scalebox{0.55}{\includegraphics*{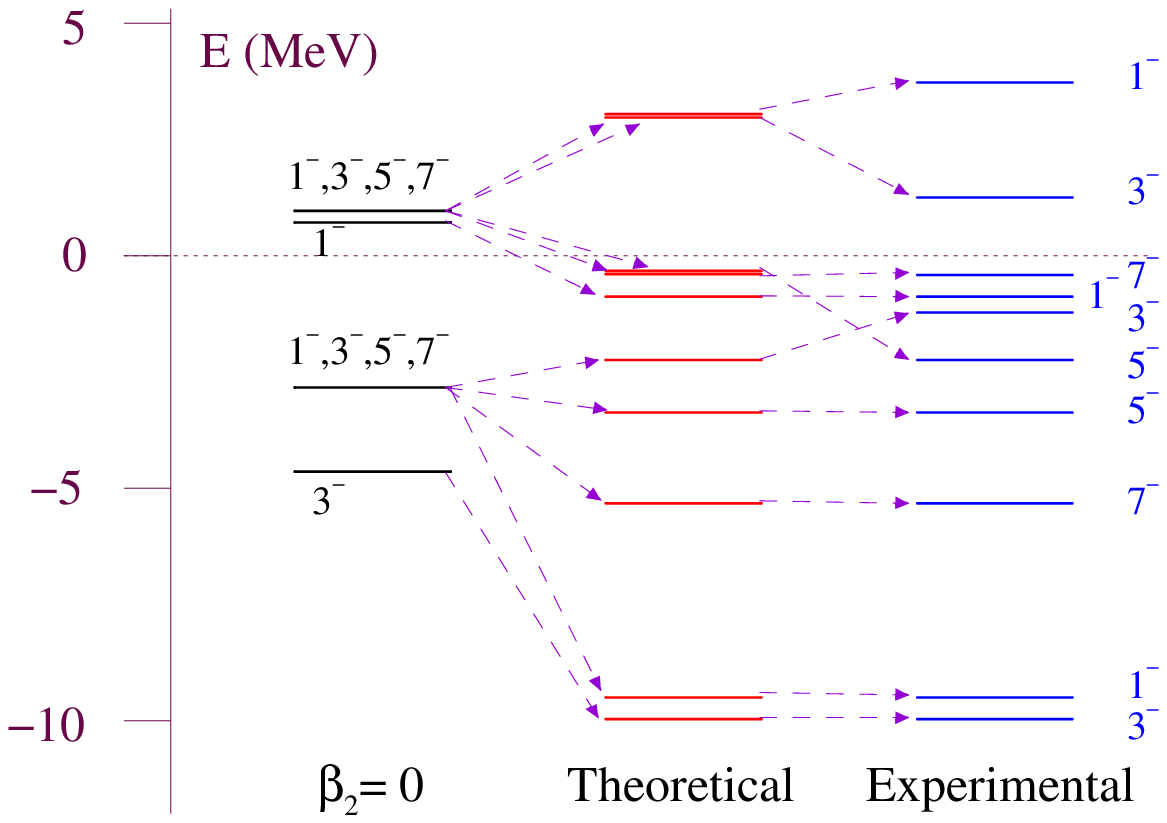}}
\vspace*{2mm}
\end{minipage} \hfill
\begin{minipage}[t]{7.0cm}
In Fig.~3.,    we show the spectra that one obtains for the $p+{}^6$He system 
when the target   (${}^6$He)   structure  is taken as three states; the $0^+$ 
(ground), and two $2^+$ states at 1.78 and 5.68 MeV respectively.           A 
quadrupole deformation was considered and first and second order deformations
allowed for each excitation.

\vspace*{5mm}
{\footnotesize {\bf FIGURE 3.}
The spectra of ${}^7$Li as determined using MCAS for the $p+{}^6$He system.}
\vspace*{2mm}
\end{minipage}\hfill
The base interaction potential required a large diffuseness consistent   with
the target having an extended neutron matter distribution. The results on the
far left came from evaluations   with   deformation  set to zero and show the 
sub-structural origins of each state in the actual      spectrum~\cite{Pi05}. 
Again use of the OPP to account for Pauli blocking of     occupied states was 
crucial in finding these results.

\section{MCAS and the optical potential}

       Assuming a local form for the elastic channel element of the potential 
matrix, the optical potential for elastic scattering is defined by
\begin{equation}
V^{opt}(r,r';E) 
= V_{1 1}(r) + 
\sum^\Gamma_{c,c'=2} V_{1 c}(r)\ 
G^{(Q)}_{cc'}(r,r';E)\
V_{c' 1}(r')\ ,
\label{opteq}
\end{equation}
where    the    second    term   is  the dynamic polarization potential (DPP)
$\Delta U(r,r';E)$.   This makes the formulated optical potential complex (if
the energy allows more than one open channel), nonlocal, and energy dependent
as $G^{(Q)}_{cc'}$ are the full Green functions referring to the       $Q\ (= 
\Gamma -1)$ excluded channels. 

    In the MCAS approach~\cite{Am03}, with ${\mbox{\boldmath $\Lambda$}}(E) = 
\left[ {\mbox{\boldmath $\eta$}}    -          {\bf G}_0^{(Q)}(E)\right]^{-1} 
- {\mbox{\boldmath $\eta$}}^{-1}$, the DPP is 
\begin{equation}
\Delta U(r,r';E) = \sum^N_{n, n' = 1} \chi_{1 n}(r) 
\left[{\mbox{\boldmath $\Lambda$}}(E) \right]_{nn'}(E) \chi_{n' 1}(r')\ .
\label{DPP}
\end{equation}
where $\chi_{n' 1}(r)$ are Bessel transforms of the from factors 
$\chi_{n' 1}(k)$.

\begin{minipage}[t]{7.0cm}
\vspace*{2mm}
\scalebox{0.5}{\includegraphics*{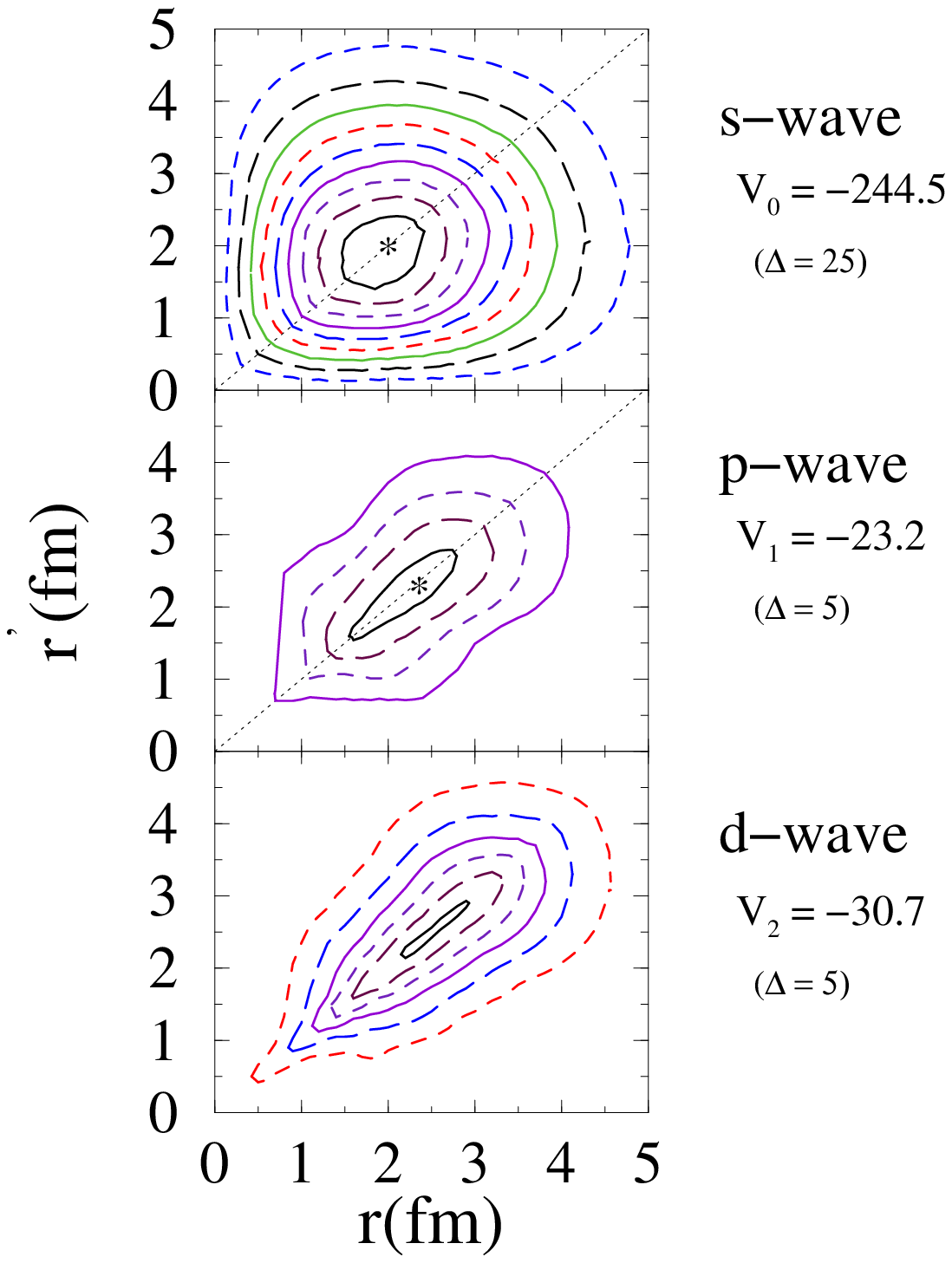}}
\vspace*{2mm}
\end{minipage} \hfill
\begin{minipage}[t]{7.0cm}
\vspace*{2mm}
The  energy-volumes of the DPP evaluated for 2.73 MeV neutrons on  ${}^{12}$C 
are plotted in Fig.~4. for three partial waves, $\ell = 0, 1, 2$. The central
depths of   each   are  indicated on the right as are the energy steps of the 
contour lines. As the energy is below the first inelastic threshold,  the DPP
is purely real.       But the coupled channel effects make that aspect of the
optical potential extremely non-local.  The DPP is also very angular momentum
dependent.

\vspace*{5mm}
{\footnotesize {\bf FIGURE 4.}
The  DPP for three partial waves for the scattering of 2.73 MeV neutrons from 
${}^{12}$C.}
\vspace*{2mm}
\end{minipage}\hfill

\section{Conclusions}

The MCAS approach to analyze (low-energy) nucleon-nucleus scattering is built
from a model structure of the interaction potentials between a nucleon    and
each of the target states taken into consideration.    All resonances (narrow
and broad) in the cross section within the selected  (positive)  energy range
can be found on a background.         With negative energies, the MCAS method
specifies the sub-threshold bound states of the compound nucleus. Inherent is
a resonance finding procedure by which all resonances      (and sub-threshold
bound) states will be found within a chosen energy range.  Also, when the OPP
is used in definition of  the Sturmians,        those functions are assured to be 
orthogonal to any single nucleon bound state that is Pauli blocked.      That
allowance for  the  influence  of  the Pauli principle was crucial in finding 
results that concur well with observation.

\bibliographystyle{aipproc}   

\bibliography{AmosK}

\end{document}